\documentclass[11pt]{article} 

\usepackage[utf8]{inputenc} 

\usepackage[margin=2.5cm]{geometry} 
\usepackage{amsfonts}
\usepackage{amsmath}
\usepackage{authblk}
\usepackage{natbib}
\usepackage{hyperref}
\usepackage{titlesec}
\titleformat*{\section}{\large\bfseries}

\setcitestyle{round}

\title{\vspace{-1.25cm} \Large 75\textsuperscript{th} Anniversary of \\ ‘Existence of Electromagnetic-Hydrodynamic Waves’}
\author{A. J. B. Russell}
\affil{School of Science \& Engineering, University of Dundee, Dundee, DD1 4HN, Scotland, U.K.}
\date{} 

\begin{document} 

\maketitle

\begin{abstract}
We have recently passed the 75th anniversary of one of the most important results in solar and space physics: Hannes Alfvén’s discovery of Alfvén waves and the Alfvén speed. To celebrate the anniversary, this article recounts some major episodes in the history of MHD waves. Following an initially cool reception, Alfvén’s ideas were propelled into the spotlight by Fermi’s work on cosmic rays, the new mystery of coronal heating and, as scientific perception of interplanetary space shifted dramatically and the space race started, detection of Alfvén waves in the solar wind. From then on, interest in MHD waves boomed, laying the foundations for modern remote observations of MHD waves in the Sun, coronal seismology and some of today’s leading theories of coronal heating and solar wind acceleration. In 1970, Alfvén received the Nobel Prize for his work in MHD, including these discoveries. The article concludes with some reflection about what the history implies about the way we do science, especially the advantages and pitfalls of idealised mathematical models. 
\end{abstract}

We have recently, on 3 October 2017, passed 75 years since Nature published the short letter that announced Hannes Alfvén’s discovery that the magnetohydrodynamic (MHD) equations imply the wave and speed that now bear his name. This article celebrates that discovery, which is one of the most important results in solar and space physics.

On the day Alfvén’s letter was published, Stalin’s and Hitler’s armies were fighting brutally at Stalingrad (now Volgograd) and Allied forces in North Africa were preparing to attack their adversaries near El Alamein. Alfvén’s single-column report (\citealt{1942Natur.150..405A}; also see \citealt{1943ArA....29Q...1A}), coming as it did in the middle of World War II, is considerably removed in time from today; and yet, its content feels instantly familiar to any modern practitioner of MHD. True, every symbol in the original Alfvén speed formula is different to those normally used today, but it is trivial to translate to the modern conventions. What is really remarkable is that Alfvén’s original model of a uniform magnetic field in the z-direction, a static equilibrium, and an incompressible perturbation of velocity in an invariant x-direction, is exactly preserved in MHD courses taught around the world today (I personally vouch for the University of Dundee’s). This can be attributed to the mathematics being both simple and unequivocal, provided one accepts the assumptions. In these respects, Alfvén’s discovery feels not unlike the landmark derivation of light waves by James Clerk Maxwell 77 years earlier \citep{1865RSPT..155..459C}.

Despite this favourable appraisal of Alfvén’s letter by modern eyes, by Alfvén’s own account it was not well received upon publication. In ‘Memoirs of a Dissident Scientist’ \citep{Memoir} he recalled that, in the first six years ‘\ldots very few people – with Lyman Spitzer and Martin Schwarzschild as the most prominent exceptions – had believed in them [MHD waves]. I got letters from colleagues who asked me whether I had not understood that this was all nonsense. If they existed, Maxwell would have described them and it is quite clear that he has not.’ Thus, while a scientific idealist would hope that a discovery like Alfvén’s would be quickly recognised, the journey towards mainstream acceptance would wait for additional factors.

In Alfvén’s own telling, the pivotal moment came when he gave a seminar about cosmic rays at Chicago in 1948. By good fortune, Enrico Fermi was in the audience and afterwards asked Alfvén to explain his ideas about MHD waves: ‘Fermi listened to what I said about them for five or ten minutes, and then he said: “Of course such waves could exist.” Fermi had such authority that if he said “of course” today, every physicist said “of course” tomorrow. Actually, he published a paper in which he explained them in such a clear way that no one could doubt their possible existence.’

The impact of social factors such as personality and reputation on the course of science will be debated for as long as there are scientists, as will the degree to which an elder scientist might, just possibly, exaggerate an anecdote decades after the events. Looking at the evidence of publication records, it is true that progress on MHD waves in the 1940s was heavily dominated by Alfvén’s circle in Stockholm, including a monopoly on the major results of Walén’s eponymous energy equipartition relation \citep{1944ArA....30....1W}, Lundquist’s experimental demonstration of Alfvén waves \citep{1949Natur.164..145L} and Herlofson’s discovery of the Alfvén, fast and slow waves for compressible media \citep{1950Natur.165.1020H}. Fermi’s paper ‘On the Origin of the Cosmic Radiation’ \citep{1949PhRv...75.1169F} was a rare international contribution in this period, not to mention one of extraordinary genius, and the fact that Fermi was a respected Nobel laureate doubtless meant the paper did not struggle for attention. It is also clear historically that international interest in MHD waves took off greatly in the following years, and the waves gained enough interest that they would be covered in various introductory plasma physics textbooks by the early sixties. Incidentally, it was at this time of increasing interest that the term ‘Alfvén wave’ came into usage, with Vincenzo Ferraro and Jim Dungey being among the first to use it \citep{1954ApJ...119..393F, 1954JGR....59..323D}.

In seeking to explain why MHD waves gained greater attention after Fermi’s intervention, a third factor besides reputation and clear argument should also be considered: the credit that MHD waves gained by being linked to a great mystery that was part of the current scientific zeitgeist, here, the origin of cosmic rays. This is an important and recurring theme in the history of Alfvén waves. Some of the attempts at application flopped, for example, Alfvén’s early attempt to explain sunspots as manifestations of MHD waves propagating up from the deep interior; this proposal was swiftly criticised by Tom Cowling in a paper that nonetheless demonstrated clear understanding and appreciation of the new wave theory \citep{1946MNRAS.106..446C}. Fortunately other applications became major successes, two of the earliest being stochastic acceleration \citep{1949PhRv...75.1169F} and Dungey’s interpretation of ultralow-frequency (ULF) waves at Earth as standing Alfvén waves \citep{1955phio.conf..229D}. From 1949 onwards, MHD waves had scientific relevance and could not be dismissed as a mere curiosity of the equations.

New applications continue to be found even today. For example, in the last decade MHD waves have become a significant focus of solar flare research beyond their role in particle acceleration. The paper by \citet{2008ApJ...675.1645F} was especially influential in opening up this area of research, 
following earlier work by Gordon Emslie, Peter Sturrock, Don Melrose and Gerhard Haerendel.  New directions in just the last two years include discoveries that chromospheric dissipation of MHD waves generated in the corona can produce flare signatures \citep{2016ApJ...818L..20R,2016ApJ...827..101K,2018ApJ...853..101R} and that large scale magnetic changes in the corona can excite sunquakes by nonlinear wave coupling \citep{2016ApJ...831...42R}. Time will tell whether the newest directions become flops or successes in their turn. 

Three quarters of a century have naturally seen very many important developments for MHD waves and for this article I must be selective, with apologies to those whose favourites and careers I miss out.  I will give an extended overview of just one area of application: the megakelvin corona and the solar wind. My reasons for choosing this one over others are that as well as being important and topical science, it had a profound impact on the history of MHD waves and it is particularly suited for showing the development of solar and space physics in the years since Alfvén’s 1942 paper.

It is fair to say that as little as two years before Alfvén’s wave paper there was no coronal heating problem. There was instead the problem of the coronal spectral lines, which became the coronal heating problem in 1941 when Bengt Edlén (building on an inspiration of Walter Grotrian’s) established that the vast majority of coronal emission lines matched transitions for highly ionised states of iron, nickel and calcium. Edlén’s work \citep{1941ArAE} implied a coronal temperature in excess of a million kelvin, which went against longstanding conceptions about the corona, but the evidence of the lines was compelling, and a high temperature would also make sense of the observed large line widths and low radial density gradient \citep{1941ArAA}. By 1945, the idea of a megakelvin corona had gained sufficient acceptance that Edlén’s results earned him the Royal Astronomical Society Gold Medal at the relatively young age of 38.

Thus, the discovery of MHD waves in 1942 coincided with a major revolution in scientific thought about stellar atmospheres, and it should surprise nobody that Alfvén threw his newly discovered waves into the ensuing fray of hypothesising about coronal heating. As early as 1947, Alfvén himself set out the basic arguments for Alfvén wave generation by granular motions, their propagation into the corona and resistive damping in the corona \citep{1947MNRAS.107..211A}; a year before Martin Schwarzschild and Ludwig Biermann individually proposed heating of the chromosphere by acoustic waves dissipating as shocks \citep{1948ApJ...107....1S,1948ZA.....25..161B}. Using a resistivity formula available to him at the time, Alfvén concluded that MHD waves produced by granular motions would heat the inner corona. Unfortunately for Alfvén, this formula gave a wrong conclusion. Six years later Cowling countered that the heating had been grossly overestimated \citep{CowlingBook} and detailed work by Jack Piddington confirmed the refutation of simple resistive damping in the corona \citep[see][]{1956MNRAS.116..314P}. Despite the misstep on resistivity, the 1947 paper convinced many of the potential for Alfvén wave heating if a suitable dissipation mechanism could be found, and the paper’s central ideas remain in play today with a couple of amendments. Our community may therefore look on the resistivity error with some gratitude, since it encouraged Alfvén to bring forward these critical ideas at a critical time.

Incidentally, the following quote from Alfvén’s 1947 paper is worth highlighting: ‘The currents associated with “granulation waves” as considered above may supply the “normal” more or less constant heating. But no doubt currents could be produced also in other ways \ldots discharges give probably an essential additive heating of the chromosphere and especially of the corona.’ The coronal heating ‘waves versus nanoflares’ question, which remains 70 years later, was there in the beginning.

As the fields of MHD waves and coronal heating took off in the fifties, another significant shift occurred in the perception of interplanetary space, and the related efforts would again prove pivotal for the progress of MHD waves. Intermittent ‘solar corpuscular radiation’ had been hypothesised for some time to explain auroral and geomagnetic activity, and alongside this one finds a few speculations about the possibility of a weaker continuous flow of ‘solar corpuscles’, particularly in connection to the 27-day recurrence of weaker geomagnetic storms and the existence of comet ion tails. Between 1951 and 1957, this matter gained serious attention as a result of Biermann’s work on comet tails, which indicated a constant flow of particles away from the Sun in all directions at speeds of the order of hundreds of km/s \citep{1951ZA.....29..274B}. Those drawn to the topic by Biermann’s results included Alfvén, who made a significant contribution by proposing that several difficulties would be overcome if the matter sweeping over the comet possessed a frozen-in magnetic field \citep{1957Tell....9...92A}. 

Around this time Sydney Chapman spotted that the high temperature of the corona combined with its high thermal conductivity and low radiation implied that hot solar plasma should extend beyond 1~AU \citep{1957SCoA....2....1C}, which raised questions as to whether the corpuscular radiation passed through this plasma or whether there was a conflict between the two theories \citep[see][]{1957Obs....77..109B}. The master stroke came from Eugene Parker, who unified the two ideas by showing that a corona with an extended temperature would not be in static equilibrium but would instead flow away from the star as a transonic solar wind (\citealt{1958ApJ...128..664P}; it is also interesting to note remarks in this direction by \citealt{1941ArAA}). The Astrophysical Journal received Parker’s solar wind paper on 2 January 1958, which was three months after Sputnik~1 became the world’s first satellite and 29 days before the USA launched their first successful satellite, Explorer~1, which discovered the Earth’s Van-Allen radiation belts. While the superpowers competed in the new arena of space flight, Parker’s paper was held up in the review process. It was finally published in November, reputedly after Subrahmanyan Chandrasekhar took an editorial decision to override the referees. Two months after the publication and almost exactly one year after Parker submitted his manuscript, Lunik~1 (1959) attempted the first in-situ measurements of interplanetary plasma. Space exploration is tricky business and instrumentation was developing rapidly, so while results from Lunik~1, Lunik~2 (1959) and Explorer~10 (1961) lent increasing support to the solar wind, science would wait for the comprehensive data from Mariner~2 (1962) before the solar wind would widely be considered as confirmed. For those interested, a comprehensive account of the discovery of the solar wind can be found in the book by \citet{1991esss.book.....H}.

Our focus returns to MHD waves at this point because as soon as high-quality measurements of the interplanetary medium were analysed, they revealed that the solar wind is full of MHD waves, especially Alfvén waves propagating away from the Sun. Extensive analyses of the interplanetary data from Mariner 2 in a series of papers by Paul Coleman Jr. \citep{1966PhRvL..17..207C,1967P&SS...15..953C,1968ApJ...153..371C}, and of Mariner 5 data in two papers by John Belcher and Leverett Davis Jr. \citep{1969JGR....74.2302B,1971JGR....76.3534B}, persuasively revealed the waves. Furthermore, the Mariner data showed that the fast solar wind required additional acceleration compared to thermal models, and MHD waves were a natural energy source to look to, especially given the compelling new evidence that they were present with large amplitudes. Parker gave wave heating prominent and favourable consideration in his 1965 review paper on the solar wind \citep{1965SSRv....4..666P}, and the search for dissipation mechanisms was reinvigorated, producing proposals such as resonant absorption, phase-mixing and turbulent dissipation. Additionally, it was not long before wave pressure was recognised as being able to contribute to a stellar wind in a manner similar to radiation pressure \citep{1971ApJ...168..509B}. 

Thus, the first quarter century in the history of MHD waves concluded with the crowning achievement of observation in the solar wind, completing the journey from controversial theory to undisputed and relevant reality.

To the present day, coronal heating and solar wind acceleration have held on to their most tantalising secrets, but MHD waves have become somewhat less mysterious. A tremendous advance has been the achievement of resolved observations of MHD waves in the Sun, which became established in the late nineties with TRACE observations of coronal slow mode and kink mode waves (this era has been described in detail by \citealt{2005LRSP....2....3N}). In the current era of SDO/AIA, such observations are abundant and give important insights into properties such as coronal magnetic field strength by application of highly refined seismological techniques. These methods are in turn based on models developed by the likes of Bernie Roberts and others who saw early on the importance of generalising Alfvén’s 1942 model to include various forms of inhomogeneity \citep[e.g.][]{1983SoPh...88..179E,2002ApJ...577..475R}. 

A second watershed moment for solar MHD wave observations occurred in 2007, when several teams succeeded in detecting propagating Alfvénic waves in the Sun, using data from the ground-based CoMP coronagraph, the Swedish Solar Telescope and the Hinode spacecraft \citep{2007Sci...317.1192T,2007SoPh..246...65L,2007Sci...318.1574D,2007Sci...318.1577O}. These findings initially attracted controversy over exactly what type of wave had been detected (kink or Alfvén or both?) but, in any interpretation, the evidence for ubiquitous waves propagating through the chromosphere and inner corona – apparently with enough energy to supply coronal heating and solar wind acceleration – was of great importance. As to whether the Sun follows the letter of Alfv\'en's early work as well as its spirit, this would soon seem to be settled by chromospheric observations of torsional Alfv\'en waves \citep{2009Sci...323.1582J}, which have the same the symmetry as the MHD wave experiments of \citet{1949Natur.164..145L}.

There is a satisfying resonance between the culmination of the first quarter century of MHD waves with the detection of Alfvén waves in the solar wind, and the breakthroughs over the last quarter century of resolved observations of MHD waves in the Sun. It finally feels like we have a grasp on two opposite ends of the puzzle and might therefore make new strides on identifying the contributions that MHD waves make to heating the corona and accelerating the solar wind. For instance, one of the major models proposes that low-frequency Alfvénic waves are weakly reflected from Alfvén speed gradients in the corona and heliosphere, and the subsequent counter-propagating waves cascade to the small length scales needed to dissipate the wave energy \citep{2005ApJS..156..265C,2014ApJ...782...81V}. In principle, this theory becomes more testable by simultaneously exploring outgoing and reflected waves in the inner corona by seismology (as recently demonstrated by Richard Morton and collaborators: \citealt{2015NatCo...6E7813M}) and outgoing and reflected waves in the interplanetary medium by in-situ measurement (as explored by Hui Li and collaborators including Belcher: \citealt{2016ApJ...824L...2L}). Optimists should, however, consider the potential impact of the blind spot between about 1.2 solar radii (upper extent of remote observations) and 9 solar radii (perihelion of the planned Parker Solar Probe). While this region may appear relatively small, it unfortunately includes the expected transonic point of the solar wind, and models are particularly sensitive to dissipation in this region. The outlook for these issues towards the century anniversary of MHD waves in 2042 therefore includes some encouraging signs, but there are grounds to believe that major questions will survive for some time yet.

On 11 December 1970, Alfvén received the Nobel Prize ‘for fundamental work and discoveries in magnetohydrodynamics with fruitful applications in different parts of plasma physics.’ The presentation speech \citep{nobelpress} made explicit reference to Alfvén’s discovery of ‘hitherto unsuspected… Alfvén waves’ and it credited him with introducing to science ‘the fundamental idea that plasma, even in space, has a magnetic field associated with it.’ It concluded by saying, perhaps with a forgivable touch of hyperbole, ‘Professor Alfvén. You have created magnetohydrodynamics.’

Considering Alfvén’s Nobel citation and the work we are celebrating, it would be easy to think that Alfvén was a prototype for the MHD theorist armed with simplified but powerful mathematical models. An irony is that Alfvén would almost certainly have resisted any such characterisation, preferring that of a plasma physicist.  In his Nobel Lecture \citep{nobellecture}, Alfvén devoted more time to his newer work on the formation of the Solar System than he did to MHD, and the remarks he did make on MHD were primarily a strong criticism of the temptations of ‘mathematically elegant’ models, warning that where measurements are available ‘it is only the plasma itself which does not understand how beautiful the theories are and absolutely refuses to obey them.’ 

There is intriguing tension between Alfvén’s remarks and the tremendous impact of his own wave model, which invites some contemplation. Some might wonder if the speech was particularly aimed at certain antagonists.
Alfv\'en made no secret of thinking that the scientific community too readily dismissed ideas that did not fit their accustomed ways of thinking, nor that he felt that this had negatively affected his own career \citep{Memoir}.
This was exacerbated by what has been called the British-Scandanavian schism \citep{2015ASSP...41..253S} and \citet{1989NYASA.571..649A} used language similar to the Nobel Lecture when he overtly described Chapman's work as `characterized more by mathematical elegance than by contact with empirical facts'.
Interwoven with this is Alfvén’s allegiance with the cause of Kristian Birkeland \citep[see][]{1984GMS....28...22D}. Birkeland’s theory of auroral field-aligned currents \citep{BirkelandBook} was proven correct shortly before Alfvén received the Nobel Prize \citep{1967CummingsDessler}, but in Birkeland’s own lifetime, 
and long thereafter, his ideas were marginalised. Birkeland was right, but his solution lay outside assumptions that others seemed to regard as truth itself (namely, that electricity cannot flow in the vacuum of space). Therefore, if Alfvén had intended his speech, in part, as a rebuke, how should the moment be interpreted? 
A major theme of classical tragedy is that heroes’ flaws are exhibited in moments of triumph. So does one see a character using the most famous stage in science to settle scores, ironically unaware of what he would most be remembered for? Or does one see a climatic moment of justice, with Birkeland recently vindicated and now celebrated through the proxy of his champion? A third dramatic perspective, to further increase the ambiguity, is that Alfvén was an arch-sceptic who often went so far as to argue against ideas he had originated when the community became enthused by them. This trait undoubtedly took him to the wrong side of several arguments, but many would say that his willingness to do this was to his credit as a scientist. So was the scepticism evident in Alfvén’s speech simply another example of a deep-rooted quality befitting a great scientist?

In any case, Alfvén’s criticisms of the dangers of allowing theory to run too far from experiment and observation, or of becoming seduced by one’s own models, are extremely sensible. 
Furthermore, the MHD community has comprehensively embraced many of the specific challenges highlighted in his speech. This list includes inhomogeneous structure (a major theme in the development of MHD wave theory from the early eighties to the present day), cross-scale coupling and magnetic reconnection. 
Other advice, particularly to pay attention to electrical currents and double layers, has been taken up by fewer MHD theorists. Happily we have many colleagues who gladly inherit the task of reminding us of these!

In considering Alfvén’s words today, we should also acknowledge that large-scale numerical simulations have radically changed the scientific landscape. Simulations typically occupy a middle ground, where physical variables can be probed more thoroughly than most observations, and where assumptions are fewer and conditions more realistic than is usually achievable analytically. Thus, as well as providing new avenues for exploration, simulations can, in principle, provide a strong bridge between theory and observations that ought to reduce the gaps between research and reality. 

On the other hand, simulations also have powerful siren qualities that belie new dangers. It is well appreciated among experienced numericists, but always worth repeating, that simulations solve only an approximation of the governing equations, which can affect conclusions in surprising ways. Cautionary tales include findings that could not be replicated using different codes, and apparent discoveries that were ultimately traced to artefacts such as a tiny but non-zero divergence of the magnetic field. A more systematic problem is that no 3-dimensional code is presently anywhere close to the same Lundquist and Reynolds numbers as the Sun, which is concerning because qualitative changes occur at critical values of these parameters: the tearing instability is a well known example and, more generally, laminar behaviour is always suspect without strong justification. Furthermore, where simulations are run with less than the full three dimensions (e.g. to maximise resolution), the loss of dimensional freedom can make the results profoundly different to the true 3D evolution. This all adds up to a sizeable potential for at least some simulations to direct attention away from, not towards, the truth. I therefore suggest that Alfvén’s warning about reality often taking no notice of a beautiful theory extends very well to beautiful simulations.

Where does this leave idealised models in solar physics and more generally? Much of this history – from Alfvén to Parker to Roberts and beyond – showcases the great value and power of well-constructed idealised mathematical models. Focusing on the essentials of a phenomenon allows profound insight and rapid progress that would not otherwise be possible. Occasionally, this opens radical new areas of inquiry and produces tremendous impacts. I would go so far as to suggest that this has been the origin of a majority of our most important theoretical concepts and terminology. The primary difficulty, seen time and time again, is the challenge of persuading colleagues of the validity and relevance of such models, and hence obtaining support for their subsequent development. In many cases this was not achieved until potential applications provided enough additional incentive, or until observations provided compelling verification. When the environment changes and a theoretical model does gain widespread acceptance, we must be careful to retain healthy scepticism, never forgetting that nature often outwits us, that convenient assumptions can be false and that cleverness is no guarantee of correct conclusions. Then begins the painstaking work of adding greater realism and advancing observations, which, we have seen with MHD waves, can be the work of many decades following an original flash of inspiration.

In 2017, Alfvén and his paper ‘Existence of Electromagnetic-Hydrodynamic Waves’ have secure legacies. The major solar physics conferences usually devote multiple sessions to the latest developments of MHD waves; the highest-impact journals regularly report new discoveries and innovations; instruments for major new spacecraft and ground facilities are designed with MHD waves in mind; and, most importantly, we continue to recruit talented students and postdocs to the field, ensuring its future. As for the history, it is a good story that continues to be told and written 75 years on.

\parskip = 0cm

\section*{Acknowledgements}

I thank two anonymous referees for their careful reviewing, insightful comments and a very good anecdote,
and the Editor for the opportunity to write this article. 
I also thank Peter Cargill, Hugh Hudson, Robertus Erdelyi and Abhishek Srivastava for interesting discussion and direction to helpful resources.
The researching and writing of this article were greatly assisted by NASA's Astrophysics Data System,
and by librarians at the University of Dundee, who obtained several of the articles cited.

\section*{Disclosure of Potential Conflicts of Interest} 
The authors declare that they have no conflicts of interest.

\bibliographystyle{spr-mp-sola}
\bibliography{AW75}

\begin{thebibliography}{56}
\ifx\bisbn     \undefined \def\bisbn  #1{ISBN #1}\fi
\ifx\binits    \undefined \def\binits#1{#1}\fi
\ifx\bauthor   \undefined \def\bauthor#1{#1}\fi
\ifx\batitle   \undefined \def\batitle#1{#1}\fi
\ifx\bjtitle   \undefined \def\bjtitle#1{\textit{#1}}\fi
\ifx\bvolume   \undefined \def\bvolume#1{\textbf{#1}}\fi
\ifx\byear     \undefined \def\byear#1{#1}\fi
\ifx\bissue    \undefined \def\bissue#1{#1}\fi
\ifx\bfpage    \undefined \def\bfpage#1{#1}\fi
\ifx\blpage    \undefined \def\blpage #1{#1}\fi
\ifx\burl      \undefined \def\burl#1{\textsf{#1}}\fi
\ifx\href      \undefined \def\href#1#2{\textsf{#2}}\fi
\ifx\betal     \undefined \def\betal{\textit{et al.}}\fi
\ifx\bctitle   \undefined \def\bctitle#1{#1}\fi
\ifx\beditor   \undefined \def\beditor#1{#1}\fi
\ifx\bbtitle   \undefined \def\bbtitle#1{\textit{#1}}\fi
\ifx\bedition  \undefined \def\bedition#1{#1}\fi
\ifx\bseriesno \undefined \def\bseriesno#1{\textbf{#1}}\fi
\ifx\blocation \undefined \def\blocation#1{#1}\fi
\ifx\bsertitle \undefined \def\bsertitle#1{\textit{#1}}\fi
\ifx\bsnm      \undefined \def\bsnm#1{#1}\fi
\ifx\bsuffix   \undefined \def\bsuffix#1{#1}\fi
\ifx\bparticle \undefined \def\bparticle#1{#1}\fi
\ifx\barticle  \undefined \def\barticle#1{}\fi
\ifx\binstitute  \undefined \def\binstitute#1{#1}\fi
\ifx\bpublisher  \undefined \def\bpublisher#1{#1}\fi
\ifx\doiurl    \undefined
  \def\doiurl#1{\href{http://dx.doi.org/#1}{\textsf{DOI}}}\fi
\ifx\arxivurl  \undefined
  \def\arxivurl#1{\href{http://arxiv.org/abs/#1}{\textsf{arXiv}}}\fi
\ifx\adsurl    \undefined
  \def\adsurl#1{\href{http://adsabs.harvard.edu/abs/#1}{\textsf{ADS}}}\fi
\ifx\botherref \undefined \def\botherref#1{}\fi
\ifx\url       \undefined \def\url#1{\textsf{#1}}\fi
\ifx\bchapter  \undefined \def\bchapter#1{}\fi
\ifx\bbook     \undefined \def\bbook#1{}\fi
\ifx\bcomment  \undefined \def\bcomment#1{#1}\fi
\ifx\oauthor   \undefined \def\oauthor#1{#1}\fi
\ifx\citeauthoryear \undefined\def \citeauthoryear#1{#1}\fi
\def\endbibitem {}
\ifx\bconflocation  \undefined \def\bconflocation#1{#1} \fi

\bibitem[\protect\citeauthoryear{{Alfv{\'e}n}}{1941}]{1941ArAA}
\begin{botherref}
\oauthor{\bsnm{{Alfv{\'e}n}}, \binits{H.}}:
1941,
{On the Solar Corona}.
\textit{Arkiv f{\"o}r Matematik, Astronomi och Fysik}
\textbf{27A}(25).
\end{botherref}
\endbibitem

\bibitem[\protect\citeauthoryear{{Alfv{\'e}n}}{1942}]{1942Natur.150..405A}
\begin{barticle}
\bauthor{\bsnm{{Alfv{\'e}n}}, \binits{H.}}:
\byear{1942},
\batitle{{Existence of Electromagnetic-Hydrodynamic Waves}}.
\bjtitle{\nat}
\bvolume{150},
\bfpage{405}.
\doiurl{10.1038/150405d0}.
\adsurl{1942Natur.150..405A}.
\end{barticle}
\endbibitem

\bibitem[\protect\citeauthoryear{{Alfv{\'e}n}}{1943}]{1943ArA....29Q...1A}
\begin{botherref}
\oauthor{\bsnm{{Alfv{\'e}n}}, \binits{H.}}:
1943,
{On the Existence of Electromagnetic-Hydrodynamic Waves}.
\textit{Arkiv f{\"o}r Matematik, Astronomi och Fysik}
\textbf{29B}(2).
\adsurl{1943ArA....29Q...1A}.
\end{botherref}
\endbibitem

\bibitem[\protect\citeauthoryear{{Alfv{\'e}n}}{1947}]{1947MNRAS.107..211A}
\begin{barticle}
\bauthor{\bsnm{{Alfv{\'e}n}}, \binits{H.}}:
\byear{1947},
\batitle{{Magneto hydrodynamic waves, and the heating of the solar corona}}.
\bjtitle{\mnras}
\bvolume{107},
\bfpage{211}.
\doiurl{10.1093/mnras/107.2.211}.
\adsurl{1947MNRAS.107..211A}.
\end{barticle}
\endbibitem

\bibitem[\protect\citeauthoryear{{Alfv{\'e}n}}{1957}]{1957Tell....9...92A}
\begin{botherref}
\oauthor{\bsnm{{Alfv{\'e}n}}, \binits{H.}}:
1957,
{On the theory of comet tails.}
\textit{Tellus}
\textbf{9}.
\adsurl{1957Tell....9...92A}.
\end{botherref}
\endbibitem

\bibitem[\protect\citeauthoryear{Alfv{\'e}n}{1970}]{nobellecture}
\begin{botherref}
\oauthor{\bsnm{Alfv{\'e}n}, \binits{H.}}:
1970,
\textit{{Nobel Lecture: Plasma Physics, Space Research and the Origin of the
  Solar System}},
Nobel Foundation, available online at \\
  https://www.nobelprize.org/nobel{\_}prizes/physics/laureates/1970/Alfv{\'e}n-lecture.html.
\end{botherref}
\endbibitem

\bibitem[\protect\citeauthoryear{{Alfv{\'e}n}}{1988}]{Memoir}
\begin{barticle}
\bauthor{\bsnm{{Alfv{\'e}n}}, \binits{H.}}:
\byear{1988},
\batitle{{Memoirs of a Dissident Scientist}}.
\bjtitle{American Scientist}
\bvolume{76},
\bfpage{249}.
\end{barticle}
\endbibitem

\bibitem[\protect\citeauthoryear{{Alfv{\'e}n}}{1989}]{1989NYASA.571..649A}
\begin{barticle}
\bauthor{\bsnm{{Alfv{\'e}n}}, \binits{H.}}:
\byear{1989},
\batitle{{Three revolutions in cosmical science from the telescope to the
  Sputnik}}.
\bjtitle{Annals of the New York Academy of Sciences}
\bvolume{571},
\bfpage{649}.
\doiurl{10.1111/j.1749-6632.1989.tb50551.x}.
\adsurl{1989NYASA.571..649A}.
\end{barticle}
\endbibitem

\bibitem[\protect\citeauthoryear{{Belcher}}{1971}]{1971ApJ...168..509B}
\begin{barticle}
\bauthor{\bsnm{{Belcher}}, \binits{J.W.}}:
\byear{1971},
\batitle{{Alf{\'e}nic Wave Pressures and the Solar Wind}}.
\bjtitle{\apj}
\bvolume{168},
\bfpage{509}.
\doiurl{10.1086/151105}.
\adsurl{1971ApJ...168..509B}.
\end{barticle}
\endbibitem

\bibitem[\protect\citeauthoryear{{Belcher} and
  {Davis}}{1971}]{1971JGR....76.3534B}
\begin{barticle}
\bauthor{\bsnm{{Belcher}}, \binits{J.W.}},
\bauthor{\bsnm{{Davis}}, \binits{L.} \bsuffix{Jr.}}:
\byear{1971},
\batitle{{Large-amplitude Alfv{\'e}n waves in the interplanetary medium, 2}}.
\bjtitle{\jgr}
\bvolume{76},
\bfpage{3534}.
\doiurl{10.1029/JA076i016p03534}.
\adsurl{1971JGR....76.3534B}.
\end{barticle}
\endbibitem

\bibitem[\protect\citeauthoryear{{Belcher}, {Davis}, and
  {Smith}}{1969}]{1969JGR....74.2302B}
\begin{barticle}
\bauthor{\bsnm{{Belcher}}, \binits{J.W.}},
\bauthor{\bsnm{{Davis}}, \binits{L.} \bsuffix{Jr.}},
\bauthor{\bsnm{{Smith}}, \binits{E.J.}}:
\byear{1969},
\batitle{{Large-amplitude Alfv{\'e}n waves in the interplanetary medium:
  Mariner 5}}.
\bjtitle{\jgr}
\bvolume{74},
\bfpage{2302}.
\doiurl{10.1029/JA074i009p02302}.
\adsurl{1969JGR....74.2302B}.
\end{barticle}
\endbibitem

\bibitem[\protect\citeauthoryear{{Biermann}}{1948}]{1948ZA.....25..161B}
\begin{barticle}
\bauthor{\bsnm{{Biermann}}, \binits{L.}}:
\byear{1948},
\batitle{{{\"U}ber die Ursache der chromosph{\"a}rischen Turbulenz und des
  UV-Exzesses der Sonnenstrahlung}}.
\bjtitle{\zap}
\bvolume{25},
\bfpage{161}.
\adsurl{1948ZA.....25..161B}.
\end{barticle}
\endbibitem

\bibitem[\protect\citeauthoryear{{Biermann}}{1951}]{1951ZA.....29..274B}
\begin{barticle}
\bauthor{\bsnm{{Biermann}}, \binits{L.}}:
\byear{1951},
\batitle{{Kometenschweife und solare Korpuskularstrahlung}}.
\bjtitle{\zap}
\bvolume{29},
\bfpage{274}.
\adsurl{1951ZA.....29..274B}.
\end{barticle}
\endbibitem

\bibitem[\protect\citeauthoryear{{Biermann}}{1957}]{1957Obs....77..109B}
\begin{barticle}
\bauthor{\bsnm{{Biermann}}, \binits{L.}}:
\byear{1957},
\batitle{{Solar corpuscular radiation and the interplanetary gas}}.
\bjtitle{The Observatory}
\bvolume{77},
\bfpage{109}.
\adsurl{1957Obs....77..109B}.
\end{barticle}
\endbibitem

\bibitem[\protect\citeauthoryear{{Birkeland}}{1908}]{BirkelandBook}
\begin{bbook}
\bauthor{\bsnm{{Birkeland}}, \binits{K.}}:
\byear{1908},
\bbtitle{{The Norwegian Aurora Polaris Expedition 1902–1903}},
\bpublisher{H. Aschehoug {\&} Co}, \blocation{Leipzip, London, New York \& Paris}
\bcomment{Available online from
  https://archive.org/details/norwegianaurorap01chririch}.
\end{bbook}
\endbibitem

\bibitem[\protect\citeauthoryear{{Chapman}}{1957}]{1957SCoA....2....1C}
\begin{barticle}
\bauthor{\bsnm{{Chapman}}, \binits{S.}}:
\byear{1957},
\batitle{{Notes on the Solar Corona and the Terrestrial Ionosphere}}.
\bjtitle{Smithsonian Contributions to Astrophysics}
\bvolume{2},
\bfpage{1}.
\adsurl{1957SCoA....2....1C}.
\end{barticle}
\endbibitem

\bibitem[\protect\citeauthoryear{{Clerk Maxwell}}{1865}]{1865RSPT..155..459C}
\begin{barticle}
\bauthor{\bsnm{{Clerk Maxwell}}, \binits{J.}}:
\byear{1865},
\batitle{{A Dynamical Theory of the Electromagnetic Field}}.
\bjtitle{Philosophical Transactions of the Royal Society of London Series I}
\bvolume{155},
\bfpage{459}.
\adsurl{1865RSPT..155..459C}.
\end{barticle}
\endbibitem

\bibitem[\protect\citeauthoryear{{Coleman}}{1966}]{1966PhRvL..17..207C}
\begin{barticle}
\bauthor{\bsnm{{Coleman}}, \binits{P.J.}}:
\byear{1966},
\batitle{{Hydromagnetic Waves in the Interplanetary Plasma}}.
\bjtitle{Physical Review Letters}
\bvolume{17},
\bfpage{207}.
\doiurl{10.1103/PhysRevLett.17.207}.
\adsurl{1966PhRvL..17..207C}.
\end{barticle}
\endbibitem

\bibitem[\protect\citeauthoryear{{Coleman}}{1967}]{1967P&SS...15..953C}
\begin{barticle}
\bauthor{\bsnm{{Coleman}}, \binits{P.J.} \bsuffix{Jr.}}:
\byear{1967},
\batitle{{Wave-like phenomena in the interplanetary plasma: Mariner 2}}.
\bjtitle{\planss}
\bvolume{15},
\bfpage{953}.
\doiurl{10.1016/0032-0633(67)90166-3}.
\adsurl{1967P\%26SS...15..953C}.
\end{barticle}
\endbibitem

\bibitem[\protect\citeauthoryear{{Coleman}}{1968}]{1968ApJ...153..371C}
\begin{barticle}
\bauthor{\bsnm{{Coleman}}, \binits{P.J.} \bsuffix{Jr.}}:
\byear{1968},
\batitle{{Turbulence, Viscosity, and Dissipation in the Solar-Wind Plasma}}.
\bjtitle{\apj}
\bvolume{153},
\bfpage{371}.
\doiurl{10.1086/149674}.
\adsurl{1968ApJ...153..371C}.
\end{barticle}
\endbibitem

\bibitem[\protect\citeauthoryear{{Cowling}}{1946}]{1946MNRAS.106..446C}
\begin{barticle}
\bauthor{\bsnm{{Cowling}}, \binits{T.G.}}:
\byear{1946},
\batitle{{Alfv{\'e}n's theory of sunspots}}.
\bjtitle{\mnras}
\bvolume{106},
\bfpage{446}.
\doiurl{10.1093/mnras/106.5.446}.
\adsurl{1946MNRAS.106..446C}.
\end{barticle}
\endbibitem

\bibitem[\protect\citeauthoryear{{Cowling}}{1953}]{CowlingBook}
\begin{bchapter}
\bauthor{\bsnm{{Cowling}}, \binits{T.G.}}:
\byear{1953},
\bctitle{8}.
\bbtitle{{The Solar System : vol. 1 The Sun, ed. by G. P. Kuiper}},
\bpublisher{Chicago University Press}, \blocation{Chicago}.
\end{bchapter}
\endbibitem

\bibitem[\protect\citeauthoryear{{Cranmer} and {van
  Ballegooijen}}{2005}]{2005ApJS..156..265C}
\begin{barticle}
\bauthor{\bsnm{{Cranmer}}, \binits{S.R.}},
\bauthor{\bsnm{{van Ballegooijen}}, \binits{A.A.}}:
\byear{2005},
\batitle{{On the Generation, Propagation, and Reflection of Alfv{\'e}n Waves
  from the Solar Photosphere to the Distant Heliosphere}}.
\bjtitle{\apjs}
\bvolume{156},
\bfpage{265}.
\doiurl{10.1086/426507}.
\adsurl{2005ApJS..156..265C}.
\end{barticle}
\endbibitem

\bibitem[\protect\citeauthoryear{{Cummings} and
  {Dessler}}{1967}]{1967CummingsDessler}
\begin{barticle}
\bauthor{\bsnm{{Cummings}}, \binits{W.D.}},
\bauthor{\bsnm{{Dessler}}, \binits{A.J.}}:
\byear{1967},
\batitle{{Field-aligned currents in the magnetosphere}}.
\bjtitle{\jgr}
\bvolume{72},
\bfpage{1007}.
\doiurl{10.1029/JZ072i003p01007}.
\adsurl{1967JGR....72.1007C}.
\end{barticle}
\endbibitem

\bibitem[\protect\citeauthoryear{{De Pontieu}
  \textit{et~al.}}{2007}]{2007Sci...318.1574D}
\begin{barticle}
\bauthor{\bsnm{{De Pontieu}}, \binits{B.}},
\bauthor{\bsnm{{McIntosh}}, \binits{S.W.}},
\bauthor{\bsnm{{Carlsson}}, \binits{M.}},
\bauthor{\bsnm{{Hansteen}}, \binits{V.H.}},
\bauthor{\bsnm{{Tarbell}}, \binits{T.D.}},
\bauthor{\bsnm{{Schrijver}}, \binits{C.J.}},
\bauthor{\bsnm{{Title}}, \binits{A.M.}},
\bauthor{\bsnm{{Shine}}, \binits{R.A.}},
\bauthor{\bsnm{{Tsuneta}}, \binits{S.}},
\bauthor{\bsnm{{Katsukawa}}, \binits{Y.}},
\bauthor{\bsnm{{Ichimoto}}, \binits{K.}},
\bauthor{\bsnm{{Suematsu}}, \binits{Y.}},
\bauthor{\bsnm{{Shimizu}}, \binits{T.}},
\bauthor{\bsnm{{Nagata}}, \binits{S.}}:
\byear{2007},
\batitle{{Chromospheric Alfv{\'e}nic Waves Strong Enough to Power the Solar
  Wind}}.
\bjtitle{Science}
\bvolume{318},
\bfpage{1574}.
\doiurl{10.1126/science.1151747}.
\adsurl{2007Sci...318.1574D}.
\end{barticle}
\endbibitem

\bibitem[\protect\citeauthoryear{{Dessler}}{1984}]{1984GMS....28...22D}
\begin{barticle}
\bauthor{\bsnm{{Dessler}}, \binits{A.J.}}:
\byear{1984},
\batitle{{The evolution of arguments regarding the existence of field-aligned
  currents}}.
\bjtitle{Washington DC American Geophysical Union Geophysical Monograph Series}
\bvolume{28},
\bfpage{22}.
\doiurl{10.1029/GM028p0022}.
\adsurl{1984GMS....28...22D}.
\end{barticle}
\endbibitem

\bibitem[\protect\citeauthoryear{{Dungey}}{1954}]{1954JGR....59..323D}
\begin{barticle}
\bauthor{\bsnm{{Dungey}}, \binits{J.W.}}:
\byear{1954},
\batitle{{The Attenuation of Alfv{\'e}n Waves}}.
\bjtitle{\jgr}
\bvolume{59},
\bfpage{323}.
\doiurl{10.1029/JZ059i003p00323}.
\adsurl{1954JGR....59..323D}.
\end{barticle}
\endbibitem

\bibitem[\protect\citeauthoryear{{Dungey}}{1955}]{1955phio.conf..229D}
\begin{bchapter}
\bauthor{\bsnm{{Dungey}}, \binits{J.W.}}:
\byear{1955},
\bctitle{{Electrodynamics of the Outer Atmosphere}}.
In: \bbtitle{{The Physics of the Ionosphere: Report of the Physical Society
  Conference Held at the Cavendish Laboratory, Cambridge, September 1954}},
\bpublisher{Physical Society}, \blocation{London},
\bfpage{229}.
\adsurl{1955phio.conf..229D}.
\end{bchapter}
\endbibitem

\bibitem[\protect\citeauthoryear{{Edl{\'e}n}}{1941}]{1941ArAE}
\begin{botherref}
\oauthor{\bsnm{{Edl{\'e}n}}, \binits{B.}}:
1941,
{An Attempt to Identify the Emission Lines in the Spectrum of the Solar
  Corona}.
\textit{Arkiv f{\"o}r Matematik, Astronomi och Fysik}
\textbf{28B}(1).
\end{botherref}
\endbibitem

\bibitem[\protect\citeauthoryear{{Edwin} and
  {Roberts}}{1983}]{1983SoPh...88..179E}
\begin{barticle}
\bauthor{\bsnm{{Edwin}}, \binits{P.M.}},
\bauthor{\bsnm{{Roberts}}, \binits{B.}}:
\byear{1983},
\batitle{{Wave propagation in a magnetic cylinder}}.
\bjtitle{\solphys}
\bvolume{88},
\bfpage{179}.
\doiurl{10.1007/BF00196186}.
\adsurl{1983SoPh...88..179E}.
\end{barticle}
\endbibitem

\bibitem[\protect\citeauthoryear{{Fermi}}{1949}]{1949PhRv...75.1169F}
\begin{barticle}
\bauthor{\bsnm{{Fermi}}, \binits{E.}}:
\byear{1949},
\batitle{{On the Origin of the Cosmic Radiation}}.
\bjtitle{Physical Review}
\bvolume{75},
\bfpage{1169}.
\doiurl{10.1103/PhysRev.75.1169}.
\adsurl{1949PhRv...75.1169F}.
\end{barticle}
\endbibitem

\bibitem[\protect\citeauthoryear{{Ferraro}}{1954}]{1954ApJ...119..393F}
\begin{barticle}
\bauthor{\bsnm{{Ferraro}}, \binits{V.C.A.}}:
\byear{1954},
\batitle{{On the Reflection and Refraction of Alfv{\'e}n Waves.}}
\bjtitle{\apj}
\bvolume{119},
\bfpage{393}.
\doiurl{10.1086/145837}.
\adsurl{1954ApJ...119..393F}.
\end{barticle}
\endbibitem

\bibitem[\protect\citeauthoryear{{Fletcher} and
  {Hudson}}{2008}]{2008ApJ...675.1645F}
\begin{barticle}
\bauthor{\bsnm{{Fletcher}}, \binits{L.}},
\bauthor{\bsnm{{Hudson}}, \binits{H.S.}}:
\byear{2008},
\batitle{{Impulsive Phase Flare Energy Transport by Large-Scale Alfv{\'e}n
  Waves and the Electron Acceleration Problem}}.
\bjtitle{\apj}
\bvolume{675},
\bfpage{1645}.
\doiurl{10.1086/527044}.
\adsurl{2008ApJ...675.1645F}.
\end{barticle}
\endbibitem

\bibitem[\protect\citeauthoryear{Gustafsson}{1970}]{nobelpress}
\begin{botherref}
\oauthor{\bsnm{Gustafsson}, \binits{T.}}:
1970,
\textit{{Nobel Prize in Physics 1970 -- Presentation Speech}},
Nobel Foundation, available online at
  https://www.nobelprize.org/nobel{\_}prizes/physics/laureates/1970/press.html.
\end{botherref}
\endbibitem

\bibitem[\protect\citeauthoryear{{Herlofson}}{1950}]{1950Natur.165.1020H}
\begin{barticle}
\bauthor{\bsnm{{Herlofson}}, \binits{N.}}:
\byear{1950},
\batitle{{Magneto-Hydrodynamic Waves in a Compressible Fluid Conductor}}.
\bjtitle{\nat}
\bvolume{165},
\bfpage{1020}.
\doiurl{10.1038/1651020a0}.
\adsurl{1950Natur.165.1020H}.
\end{barticle}
\endbibitem

\bibitem[\protect\citeauthoryear{{Hufbauer}}{1991}]{1991esss.book.....H}
\begin{bbook}
\bauthor{\bsnm{{Hufbauer}}, \binits{K.}}:
\byear{1991},
\bbtitle{{Exploring the Sun : Solar Science since Galileo}},
\bpublisher{The Johns Hopkins University Press}, \blocation{Baltimore \& London}
\adsurl{1991esss.book.....H}.
\end{bbook}
\endbibitem

\bibitem[\protect\citeauthoryear{{Jess}
  \textit{et~al.}}{2009}]{2009Sci...323.1582J}
\begin{barticle}
\bauthor{\bsnm{{Jess}}, \binits{D.B.}},
\bauthor{\bsnm{{Mathioudakis}}, \binits{M.}},
\bauthor{\bsnm{{Erd{\'e}lyi}}, \binits{R.}},
\bauthor{\bsnm{{Crockett}}, \binits{P.J.}},
\bauthor{\bsnm{{Keenan}}, \binits{F.P.}},
\bauthor{\bsnm{{Christian}}, \binits{D.J.}}:
\byear{2009},
\batitle{{Alfv{\'e}n Waves in the Lower Solar Atmosphere}}.
\bjtitle{Science}
\bvolume{323},
\bfpage{1582}.
\doiurl{10.1126/science.1168680}.
\adsurl{2009Sci...323.1582J}.
\end{barticle}
\endbibitem

\bibitem[\protect\citeauthoryear{{Kerr}
  \textit{et~al.}}{2016}]{2016ApJ...827..101K}
\begin{barticle}
\bauthor{\bsnm{{Kerr}}, \binits{G.S.}},
\bauthor{\bsnm{{Fletcher}}, \binits{L.}},
\bauthor{\bsnm{{Russell}}, \binits{A.J.B.}},
\bauthor{\bsnm{{Allred}}, \binits{J.C.}}:
\byear{2016},
\batitle{{Simulations of the Mg II k and Ca II 8542 lines from an Alfv{\'E}n
  Wave-heated Flare Chromosphere}}.
\bjtitle{\apj}
\bvolume{827},
\bfpage{101}.
\doiurl{10.3847/0004-637X/827/2/101}.
\adsurl{2016ApJ...827..101K}.
\end{barticle}
\endbibitem

\bibitem[\protect\citeauthoryear{{Li}
  \textit{et~al.}}{2016}]{2016ApJ...824L...2L}
\begin{barticle}
\bauthor{\bsnm{{Li}}, \binits{H.}},
\bauthor{\bsnm{{Wang}}, \binits{C.}},
\bauthor{\bsnm{{Belcher}}, \binits{J.W.}},
\bauthor{\bsnm{{He}}, \binits{J.}},
\bauthor{\bsnm{{Richardson}}, \binits{J.D.}}:
\byear{2016},
\batitle{{Sunward-propagating Alfv{\'e}nic Fluctuations Observed in the
  Heliosphere}}.
\bjtitle{\apjl}
\bvolume{824},
\bfpage{L2}.
\doiurl{10.3847/2041-8205/824/1/L2}.
\adsurl{2016ApJ...824L...2L}.
\end{barticle}
\endbibitem

\bibitem[\protect\citeauthoryear{{Lin}
  \textit{et~al.}}{2007}]{2007SoPh..246...65L}
\begin{barticle}
\bauthor{\bsnm{{Lin}}, \binits{Y.}},
\bauthor{\bsnm{{Engvold}}, \binits{O.}},
\bauthor{\bsnm{{Rouppe van der Voort}}, \binits{L.H.M.}},
\bauthor{\bsnm{{van Noort}}, \binits{M.}}:
\byear{2007},
\batitle{{Evidence of Traveling Waves in Filament Threads}}.
\bjtitle{\solphys}
\bvolume{246},
\bfpage{65}.
\doiurl{10.1007/s11207-007-0402-8}.
\adsurl{2007SoPh..246...65L}.
\end{barticle}
\endbibitem

\bibitem[\protect\citeauthoryear{{Lundquist}}{1949}]{1949Natur.164..145L}
\begin{barticle}
\bauthor{\bsnm{{Lundquist}}, \binits{S.}}:
\byear{1949},
\batitle{{Experimental Demonstration of Magneto-hydrodynamic Waves}}.
\bjtitle{\nat}
\bvolume{164},
\bfpage{145}.
\doiurl{10.1038/164145a0}.
\adsurl{1949Natur.164..145L}.
\end{barticle}
\endbibitem

\bibitem[\protect\citeauthoryear{{Morton}, {Tomczyk}, and
  {Pinto}}{2015}]{2015NatCo...6E7813M}
\begin{barticle}
\bauthor{\bsnm{{Morton}}, \binits{R.J.}},
\bauthor{\bsnm{{Tomczyk}}, \binits{S.}},
\bauthor{\bsnm{{Pinto}}, \binits{R.}}:
\byear{2015},
\batitle{{Investigating Alfv{\'e}nic wave propagation in coronal open-field
  regions}}.
\bjtitle{Nature Communications}
\bvolume{6},
\bfpage{7813}.
\doiurl{10.1038/ncomms8813}.
\adsurl{2015NatCo...6E7813M}.
\end{barticle}
\endbibitem

\bibitem[\protect\citeauthoryear{{Nakariakov} and
  {Verwichte}}{2005}]{2005LRSP....2....3N}
\begin{barticle}
\bauthor{\bsnm{{Nakariakov}}, \binits{V.M.}},
\bauthor{\bsnm{{Verwichte}}, \binits{E.}}:
\byear{2005},
\batitle{{Coronal Waves and Oscillations}}.
\bjtitle{Living Reviews in Solar Physics}
\bvolume{2},
\bfpage{3}.
\doiurl{10.12942/lrsp-2005-3}.
\adsurl{2005LRSP....2....3N}.
\end{barticle}
\endbibitem

\bibitem[\protect\citeauthoryear{{Okamoto}
  \textit{et~al.}}{2007}]{2007Sci...318.1577O}
\begin{barticle}
\bauthor{\bsnm{{Okamoto}}, \binits{T.J.}},
\bauthor{\bsnm{{Tsuneta}}, \binits{S.}},
\bauthor{\bsnm{{Berger}}, \binits{T.E.}},
\bauthor{\bsnm{{Ichimoto}}, \binits{K.}},
\bauthor{\bsnm{{Katsukawa}}, \binits{Y.}},
\bauthor{\bsnm{{Lites}}, \binits{B.W.}},
\bauthor{\bsnm{{Nagata}}, \binits{S.}},
\bauthor{\bsnm{{Shibata}}, \binits{K.}},
\bauthor{\bsnm{{Shimizu}}, \binits{T.}},
\bauthor{\bsnm{{Shine}}, \binits{R.A.}},
\bauthor{\bsnm{{Suematsu}}, \binits{Y.}},
\bauthor{\bsnm{{Tarbell}}, \binits{T.D.}},
\bauthor{\bsnm{{Title}}, \binits{A.M.}}:
\byear{2007},
\batitle{{Coronal Transverse Magnetohydrodynamic Waves in a Solar Prominence}}.
\bjtitle{Science}
\bvolume{318},
\bfpage{1577}.
\doiurl{10.1126/science.1145447}.
\adsurl{2007Sci...318.1577O}.
\end{barticle}
\endbibitem

\bibitem[\protect\citeauthoryear{{Parker}}{1958}]{1958ApJ...128..664P}
\begin{barticle}
\bauthor{\bsnm{{Parker}}, \binits{E.N.}}:
\byear{1958},
\batitle{{Dynamics of the Interplanetary Gas and Magnetic Fields.}}
\bjtitle{\apj}
\bvolume{128},
\bfpage{664}.
\doiurl{10.1086/146579}.
\adsurl{1958ApJ...128..664P}.
\end{barticle}
\endbibitem

\bibitem[\protect\citeauthoryear{{Parker}}{1965}]{1965SSRv....4..666P}
\begin{barticle}
\bauthor{\bsnm{{Parker}}, \binits{E.N.}}:
\byear{1965},
\batitle{{Dynamical Theory of the Solar Wind}}.
\bjtitle{\ssr}
\bvolume{4},
\bfpage{666}.
\doiurl{10.1007/BF00216273}.
\adsurl{1965SSRv....4..666P}.
\end{barticle}
\endbibitem

\bibitem[\protect\citeauthoryear{{Piddington}}{1956}]{1956MNRAS.116..314P}
\begin{barticle}
\bauthor{\bsnm{{Piddington}}, \binits{J.H.}}:
\byear{1956},
\batitle{{Solar atmospheric heating by hydromagnetic waves}}.
\bjtitle{\mnras}
\bvolume{116},
\bfpage{314}.
\doiurl{10.1093/mnras/116.3.314}.
\adsurl{1956MNRAS.116..314P}.
\end{barticle}
\endbibitem

\bibitem[\protect\citeauthoryear{{Reep} and
  {Russell}}{2016}]{2016ApJ...818L..20R}
\begin{barticle}
\bauthor{\bsnm{{Reep}}, \binits{J.W.}},
\bauthor{\bsnm{{Russell}}, \binits{A.J.B.}}:
\byear{2016},
\batitle{{Alfv{\'e}nic Wave Heating of the Upper Chromosphere in Flares}}.
\bjtitle{\apjl}
\bvolume{818},
\bfpage{L20}.
\doiurl{10.3847/2041-8205/818/1/L20}.
\adsurl{2016ApJ...818L..20R}.
\end{barticle}
\endbibitem

\bibitem[\protect\citeauthoryear{{Reep}
  \textit{et~al.}}{2018}]{2018ApJ...853..101R}
\begin{barticle}
\bauthor{\bsnm{{Reep}}, \binits{J.W.}},
\bauthor{\bsnm{{Russell}}, \binits{A.J.B.}},
\bauthor{\bsnm{{Tarr}}, \binits{L.A.}},
\bauthor{\bsnm{{Leake}}, \binits{J.E.}}:
\byear{2018},
\batitle{{A Hydrodynamic Model of Alfv{\'e}nic Wave Heating in a Coronal Loop
  and Its Chromospheric Footpoints}}.
\bjtitle{\apj}
\bvolume{853},
\bfpage{101}.
\doiurl{10.3847/1538-4357/aaa2fe}.
\adsurl{2018ApJ...853..101R}.
\end{barticle}
\endbibitem

\bibitem[\protect\citeauthoryear{{Ruderman} and
  {Roberts}}{2002}]{2002ApJ...577..475R}
\begin{barticle}
\bauthor{\bsnm{{Ruderman}}, \binits{M.S.}},
\bauthor{\bsnm{{Roberts}}, \binits{B.}}:
\byear{2002},
\batitle{{The Damping of Coronal Loop Oscillations}}.
\bjtitle{\apj}
\bvolume{577},
\bfpage{475}.
\doiurl{10.1086/342130}.
\adsurl{2002ApJ...577..475R}.
\end{barticle}
\endbibitem

\bibitem[\protect\citeauthoryear{{Russell}
  \textit{et~al.}}{2016}]{2016ApJ...831...42R}
\begin{barticle}
\bauthor{\bsnm{{Russell}}, \binits{A.J.B.}},
\bauthor{\bsnm{{Mooney}}, \binits{M.K.}},
\bauthor{\bsnm{{Leake}}, \binits{J.E.}},
\bauthor{\bsnm{{Hudson}}, \binits{H.S.}}:
\byear{2016},
\batitle{{Sunquake Generation by Coronal Magnetic Restructuring}}.
\bjtitle{\apj}
\bvolume{831},
\bfpage{42}.
\doiurl{10.3847/0004-637X/831/1/42}.
\adsurl{2016ApJ...831...42R}.
\end{barticle}
\endbibitem

\bibitem[\protect\citeauthoryear{{Schwarzschild}}{1948}]{1948ApJ...107....1S}
\begin{barticle}
\bauthor{\bsnm{{Schwarzschild}}, \binits{M.}}:
\byear{1948},
\batitle{{On Noise Arising from the Solar Granulation.}}
\bjtitle{\apj}
\bvolume{107},
\bfpage{1}.
\doiurl{10.1086/144983}.
\adsurl{1948ApJ...107....1S}.
\end{barticle}
\endbibitem

\bibitem[\protect\citeauthoryear{{Southwood}}{2015}]{2015ASSP...41..253S}
\begin{bchapter}
\bauthor{\bsnm{{Southwood}}, \binits{D.}}:
\byear{2015},
\bctitle{{From the Carrington Storm to the Dungey Magnetosphere}}.
In: \beditor{\bsnm{{Southwood}}, \binits{D.}},
\beditor{\bsnm{{Cowley FRS}}, \binits{S.W.H.}},
\beditor{\bsnm{{Mitton}}, \binits{S.}} (eds.)
\bbtitle{Magnetospheric Plasma Physics: The Impact of Jim Dungey's Research},
\bsertitle{Astrophysics and Space Science Proceedings}
\bseriesno{41},
\bfpage{253}.
\adsurl{2015ASSP...41..253S}.
\end{bchapter}
\endbibitem

\bibitem[\protect\citeauthoryear{{Tomczyk}
  \textit{et~al.}}{2007}]{2007Sci...317.1192T}
\begin{barticle}
\bauthor{\bsnm{{Tomczyk}}, \binits{S.}},
\bauthor{\bsnm{{McIntosh}}, \binits{S.W.}},
\bauthor{\bsnm{{Keil}}, \binits{S.L.}},
\bauthor{\bsnm{{Judge}}, \binits{P.G.}},
\bauthor{\bsnm{{Schad}}, \binits{T.}},
\bauthor{\bsnm{{Seeley}}, \binits{D.H.}},
\bauthor{\bsnm{{Edmondson}}, \binits{J.}}:
\byear{2007},
\batitle{{Alfv{\'e}n Waves in the Solar Corona}}.
\bjtitle{Science}
\bvolume{317},
\bfpage{1192}.
\doiurl{10.1126/science.1143304}.
\adsurl{2007Sci...317.1192T}.
\end{barticle}
\endbibitem

\bibitem[\protect\citeauthoryear{{van der Holst}
  \textit{et~al.}}{2014}]{2014ApJ...782...81V}
\begin{barticle}
\bauthor{\bsnm{{van der Holst}}, \binits{B.}},
\bauthor{\bsnm{{Sokolov}}, \binits{I.V.}},
\bauthor{\bsnm{{Meng}}, \binits{X.}},
\bauthor{\bsnm{{Jin}}, \binits{M.}},
\bauthor{\bsnm{{Manchester}}, \binits{W.B.} \bsuffix{IV}},
\bauthor{\bsnm{{T{\'o}th}}, \binits{G.}},
\bauthor{\bsnm{{Gombosi}}, \binits{T.I.}}:
\byear{2014},
\batitle{{Alfv{\'e}n Wave Solar Model (AWSoM): Coronal Heating}}.
\bjtitle{\apj}
\bvolume{782},
\bfpage{81}.
\doiurl{10.1088/0004-637X/782/2/81}.
\adsurl{2014ApJ...782...81V}.
\end{barticle}
\endbibitem

\bibitem[\protect\citeauthoryear{{Wal{\'e}n}}{1944}]{1944ArA....30....1W}
\begin{barticle}
\bauthor{\bsnm{{Wal{\'e}n}}, \binits{C.}}:
\byear{1944},
\batitle{{On the Theory of Sunspots}}.
\bjtitle{Arkiv f{\"o}r Matematik, Astronomi och Fysik}
\bvolume{30},
\bfpage{1}.
\adsurl{1944ArA....30....1W}.
\end{barticle}
\endbibitem

\end{thebibliography}

\end{document}